\documentclass[twocolumn,aps,showpacs,preprintnumbers,amsmath,amssymb]{revtex4}

\usepackage{graphicx}

\begin{document}

\title{Spatial Correlation Functions of one-dimensional Bose gases at Equilibrium}
\author{N.P. Proukakis}
\address{School of Mathematics and Statistics, University of Newcastle, Merz Court, 
Newcastle NE1 7RU, United Kingdom} 

\begin{abstract}

The dependence of the three lowest order spatial correlation functions of a harmonically
confined Bose gas on temperature and interaction strength is presented at equilibrium.
Our analysis is based on
a stochastic Langevin equation for the order parameter of a weakly-interacting gas.
Comparison of the predicted 
first order correlation functions to those of  
appropriate mean field theories demonstrates the potentially crucial role
of density fluctuations on the equilibrium coherence length. Furthermore,
the change in both coherence length and shape of the correlation function, 
from gaussian to exponential, with increasing temperature
is quantified.
Moreover, the presented results for higher order correlation functions are shown to be in agreeement with
existing predictions.
Appropriate consideration of density-density correlations is shown to facilitate a
precise determination of quasi-condensate density profiles, 
providing an alternative approach to the bimodal
density fits typically used experimentally.

\end{abstract}


\maketitle

\section{Introduction}

The appearance of coherence in a system,
which is central to our understanding of laser and matter wave physics,
depends on the interplay between various parameters,
such as temperature, interaction strength, confinement, and dimensionality.
In particular, in the recently achieved confined weakly-interacting
quasi-one-dimensional Bose gases in harmonic traps
\cite{Low_D_MIT,Low_D_ENS,Low_D_Exp_1,Low_D_Exp_2,Low_D_Exp_3,Low_D_Exp_4}, dipole traps \cite{Box_BEC,Oberthaler_3} and atom chips \cite{Atom_Chip_Tubingen,Atom_Chip_Munich,Atom_Chip_MIT,Atom_Chip_Schmiedmayer,Atom_Chip_Imperial,Atom_Chip_JILA,Atom_Chip_Stanford,Atom_Chip_ENS,Atom_Chip_Brisbane,Atom_Chip_New,Real_1D_2} coherence can only be maintained across the entire spatial extent of the system
at a sufficiently low temperature, which is however much lower than the `critical temperature'
for the onset of quantum degeneracy.
At intermediate temperatures, 
long wavelength fluctuations in the phase, restrict the
coherence to smaller regions
\cite{popov,Low_D_Theory_1,Low_D_Theory_2a,Low_D_Theory_2b,Low_D_Theory_2c,Low_D_Theory_2d,Low_D_Theory_3a,Low_D_Theory_3b,Low_D_Theory_3c,Low_D_Theory_3d,Low_D_Theory_3e,Low_D_Theory_3f,Low_D_Theory_3g}. This effect has already been observed experimentally
in very elongated three-dimensional Bose-Einstein Condensates 
\cite{Low_D_Exp_1,Low_D_Exp_2,Low_D_Exp_3,Low_D_Exp_4,Low_D_Hannover_PRA,Low_D_ENS_EPJD,Low_D_ENS_Density_Fluctuations}.

In his seminal work \cite{Glauber}, Glauber characterised the coherence of a system by means of a set of 
normalised correlation functions. Typically, the few `lowest order' correlation functions
are enough to characterise the coherence of a system \cite{QO_Book_1,QO_Book_2}.
Perhaps the most important correlation function is the 
first order correlation
function, referred to in condensed matter literature as the
off-diagonal one-body density matrix. 
In an atomic gas above the degeneracy temperature, this quantity decays rapidly to zero, on a 
scale comparable to the atomic de Broglie wavelength (see, e.g., \cite{Tannoudji}).
In the opposite extreme, when this quantity tends to a nonzero plateau, the system 
becomes coherent and is said to
contain a Bose-Einstein Condensate (BEC).
In degenerate one-dimensional (1D) gases, however, an intermediate state exists,
in which the off-diagonal one-body density matrix does decay to zero within the system size, but 
at a much slower rate. 
In this regime, the system is said to contain a quasi-condensate \cite{popov}.

The phase coherence properties of such trapped quasi-condensates were
first investigated by Petrov et al. \cite{Low_D_Theory_1} in the low temperature limit,
where quasi-condensate depletion can be treated as negligible.
A number of alternative mean field approaches have appeared
in the literature since \cite{Low_D_Theory_2a,Low_D_Theory_2b,Low_D_Theory_2c,Low_D_Theory_2d,Low_D_Theory_3a,Low_D_Theory_3b,Low_D_Theory_3c,Low_D_Theory_3d,Low_D_Theory_3e,Low_D_Theory_3f,Low_D_Theory_3g,Low_D_Lee,Low_D_Walser}. In particular, the theory of Andersen et al. \cite{Low_D_Theory_2a} treats both phase and
density fluctuations in a self-consistent manner, 
thus providing the natural extension of the theory of Petrov et al. 
\cite{Low_D_Theory_1} to finite temperatures.
Using this theory, we constructed the universal one-dimensional (1D) phase diagram in the
weakly-interacting regime \cite{Low_D_Theory_2c}.
We also showed that inclusion of density fluctuations can potentially lead to a significant
decrease in the equilibrium coherence length of a trapped quasi-condensate  
\cite{Low_D_Theory_2d}.
Although initial experiments performed to address this question found no such effect within their experimental
resolution \cite{Low_D_Exp_2,Low_D_Exp_3,Low_D_Exp_4,Low_D_Theory_1b}, we argued that the
apparent insensitivity of the (suitably scaled) coherence length on density fluctuations only arises in
 the regime $T_{\phi} \ll T_{c}$, in which all experiments to date have been performed,
where $T_{\phi}$ is the characteristic temperature for phase fluctuations, and $T_{c}$ is the
`critical temperature' for the onset of (quasi)condensation.
In fact, despite a relatively weak
coupling between density and phase fluctuations
in most experiments,
the first evidence of such coupling
may have actually been recently observed \cite{Low_D_ENS_EPJD}.
The optimum conditions for the unequivocal experimental demonstration of such coupling
between density and phase fluctuations was laid out in
 \cite{Low_D_Theory_2d}.

In this paper, we present results for the three lowest order spatial correlation functions 
at equilibrium, based on the
stochastic Langevin treatment of Stoof 
\cite{Stoof_Noisy_1,Stoof_Noisy_2}.
Where appropriate, these results are compared and contrasted to predictions of various
mean field theories \cite{Low_D_Theory_1,Low_D_Theory_2a,Low_D_Theory_3a,Davis_Class_Trap,Kheruntsyan_g2_2}.
Firstly, we discuss the dependence of a suitably-defined
 equilibrium coherence length on temperature.
Predictions of the present treatment are shown to be in full agreement with the mean field theory of Andersen et al. \cite{Low_D_Theory_2a}, which includes density fluctuations in an ab initio manner, thus complementing  and extending
earlier related work \cite{Low_D_Theory_2b}.
Furthermore, the present work provides additional evidence to support our earlier claim \cite{Low_D_Theory_2d} (based on mean field theory) that the coherence length is indeed sensitive to density fluctuations and quasi-condensate depletion in realistic parameter regimes.
The spatial variation of the off-diagonal first order correlation function with distance $z$ from the trap centre ($z=0$) is
evaluated in two different ways, corresponding experimentally to the particular atomic
interference experiment that could be performed: 
In the first approach, atoms at the centre are numerically `interfered' with atoms
at a distance $z$ from the trap centre, whereas the other approach  `interferes' atoms located
symmetrically about the trap centre, i.e., at points $\pm z/2$.
Good agreement is found between these two approaches within the quasi-condensate region.
The interplay between quasi-condensation and `true' condensation is known to be manifested, not only in
the change of the scaled coherence length, but also
 in the shape of the
first order correlation function, which changes smoothly from exponential to gaussian \cite{Low_D_Theory_3d}.
Analysing the correlation functions obtained by the stochastic approach at different temperatures,
we model the interplay between such behaviour by two parameters, the first one controlling
the shape of the function, and the second representing an appropriately-defined 
coherence length, which decreases uniformly with increasing temperature.

Higher order
density-density correlation functions contain crucial information regarding the system's coherence, and
determination of such correlations (indirectly via measurement of the collisional interaction energy \cite{g2a,g2b,g2c},
or the three-body inelastic loss rate \cite{kagan,g3}) has played a key role in 
unequivocally demonstrating the 
experimental observation of BEC in ultracold 3D bosonic atomic gases. 
Moreover, second order correlation functions have been well-studied in the context of photon statistics
in the Hanbury-Brown-Twiss experiment \cite{HBT_Exp}, to discriminate between thermal and coherent light fields \cite{QO_Book_1,QO_Book_2}, and a
similar technique was recently used to experimentally 
determine the phase coherence length of a trapped elongated quasi-condensate \cite{Low_D_Exp_3,Low_D_Hannover_PRA}.
Furthermore, spatial two-body correlations have been observed in 
expanding atom clouds \cite{Correlations_Folling,Correlations_Schellekens}
and atoms produced from molecular breakup \cite{Correlations_Greiner},
whereas temporal correlations were measured in continuously outcoupled atom lasers \cite{Correlations_AtomLaser}.
This paper additionally investigates in detail the dependence of the
second and third order spatial correlation functions in confined weakly-interacting 1D Bose gases
on temperature and interaction strength.
The second order correlation function has already been discussed in the weakly-interacting regime in \cite{Davis_Class_Trap},
with the crossover to the strongly-interacting regime presented in \cite{Kheruntsyan_g2_1,Kheruntsyan_g2_2},
and our results reveal good agreement with such treatments in the appropriate regimes.

Density-density correlations are also used to directly determine the amount of
quasi-condensation present in a system at any given instant.
Although already applied to study quasi-condensate 
growth on an atom chip \cite{Dimple_Growth_PRA},
the details of this technique were not explicitly presented in our earlier work.
In our opinion, this approach provides the most direct method that can be
experimentally applied
to obtain an unequivocal determination of quasi-condensate density profiles.
In addition to providing an alternative to the conventional bimodal density fits, 
the advantage of experimentally obtaining the quasi-condensate density by the proposed method
is that it is not subject to any potential limitations 
of the particular theory used to analyze the experimental data.
This technique can be readily applied to existing experiments which performed in situ
measurements of density fluctuations of quasi-1D Bose gases at equilibrium 
\cite{Low_D_ENS_Density_Fluctuations}.

This paper is structured as follows: Sec. II briefly outlines the stochastic approach and 
other mean field theories to which the results are compared (see also Appendix A).
Sec. III discusses the first order correlation function, focusing on the dependence of an 
appropriately defined coherence length
on temperature (Sec. III A), and on method of evaluation (Sec. III B);
moreover, Sec. III C analyses the crossover in the shape of the first order correlation function
 with decreasing temperature.
Sec. IV discusses the dependence of 
higher-order density-density correlation functions on position and temperature (Sec. IV A) and 
effective interaction strength (Sec. IV B). Sec. V expounds how measurement of
density-density correlations can provide an unequivocal
determination of quasi-condensate density, which is demonstrated by means of suitable examples.
Finally, Sec. VI summarises the main results of this paper.

\section{Low-dimensional Theories}

We start by reviewing the stochastic approach which will be used to evaluate all
correlation functions presented in this work, and also briefly outlining the particular mean field theories against
which our results will be tested.

\subsection{Stochastic Langevin Approach}

The stochastic Langevin theory of Stoof \cite{Stoof_Noisy_1,Stoof_Noisy_2} is a
non-equilibrium approach, in which the system dynamics is obtained via a Langevin equation governing the evolution 
of the quasi-condensate order parameter $\Phi(z,t)$, given by
\cite{Stoof_Noisy_2}
\begin{eqnarray} i \hbar \frac{ \partial \Phi(z,t) }{ \partial t} 
& = & \Bigg[ - \frac {\hbar^{2} \nabla^{2} }{2m} + V^{\rm ext}(z) - \mu 
- iR(z,t) 
\nonumber \\ & & 
+ g |\Phi(z,t)|^{2} \Bigg] \Phi(z,t) + \eta(z,t)\;.
\label{lang}
\end{eqnarray}
In the present work, this equation is applied to study the growth of a 1D degenerate atomic gas
in contact with a thermal cloud which acts as its heat bath. The numerical implementation
of this scheme, along with further details, are discussed in Refs. \cite{Stoof_Noisy_2,Stoof_Noisy_3},
while alternative but related approaches can be found in Refs.
\cite{Stoch_Gardiner_1,Stoch_Gardiner_2,Davis_Class_PRL,Davis_Class_Trap,Stoch_Castin}.

In addition to the usual kinetic, potential and interaction terms appearing within the square brackets of Eq. (1),
Stoof's Langevin equation contains the contribution
$iR(z,t)$, which describes
pumping of the one-dimensional gas from the surrounding thermal reservoir. 
In the classical approximation imposed here for numerical simplicity, whereby the non-condensed 1D atomic cloud relaxes to the
`classical' value $N(\varepsilon) = \left[ \beta (\varepsilon - \mu) \right]^{-1}$, the above pumping term obeys
\begin{eqnarray}
iR(z,t)&=&-{\beta\over4}\hbar\Sigma^{\rm K}(z)
\nonumber \\&&
\hspace{-2cm}
\times
\left(
-{\hbar^2\nabla^2\over2m}+V^{\rm ext}(z)-\mu+g|\Phi(z,t)|^2
\right)\;.
\end{eqnarray}
The term $\eta(z,t)$ corresponds to associated noise arising from the random nature of collisions occuring in the system.
This noise term has Gaussian correlations of the form
\begin{eqnarray}
\langle
\eta^*(z,t)\eta(z^\prime,t^{\prime})
\rangle
&=&{i\hbar^2\over2}\Sigma^{\rm K}(z)\delta(z-z^\prime)\delta(t-t^{\prime})\;,
\end{eqnarray}
where $\langle...\rangle$ denotes averaging over the realizations 
of the noise. 
The quantities of Eqs. (2) and (3) depend on the one-dimensional Keldysh
self-energy, $\hbar\Sigma^{K}(z)$, which accounts both
 for collisions that send a thermal atom into the quasi-condensate,
and those which promote a quasi-condensate atom into the thermal cloud. The formulation of this theory
ensures that the trapped gas relaxes to the correct equilibrium, in accordance with the
fluctuation-dissipation theorem. The quantity $\mu$ appearing in Eqs. (1)-(2) describes the chemical potential
of the system, whereas the external potential $V^{\rm ext}(z)$ is 
treated as harmonic throughout this work,
i.e. $V^{\rm ext}(z) = m \omega_{z}^{2} z^{2}/2$, where $\omega_{z}$ the corresponding confining frequency. 

The quantity $\Phi(z,t)$ contains information
about both mean field and fluctuations around it. It therefore implicitly includes both density and 
phase fluctuations, and acts as an approximation to the
Bose field operator $\hat{\Psi}(z,t)$.
Predictions of this theory are based upon averaging the Langevin field $\Phi(z,t)$
over different noise realizations.
In particular, this theory enables an accurate determination of both diagonal and off-diagonal,
spatial and temporal correlation functions of any order
at any spatial and temporal coordinate. This is achieved by
 suitable numerical autocorrelation measurements
of the `order parameter' $\Phi(z,t)$, upon making the following identification
\begin{equation}
\langle \hat{\Psi}^{\dag}(z,t) \hat{\Psi}(z',t') \rangle 
\rightarrow \langle \Phi^{*}(z,t) \Phi(z',t') \rangle\;.
\end{equation}

In this paper we restrict our analysis to spatial correlations at equilibrium.
To study these, we first evolve the system for a sufficiently long time, $t_{eq}$, such that it relaxes
to the correct equilibrium, before performing the desired same-time autocorrelation measurements.
The first order normalised off-diagonal correlation function
$g^{(1)}(0,z)$ associated with fluctuations in the quasi-condensate phase is thus calculated 
with respect to the centre of the trap via
\begin{equation}
g^{(1)}(0,z;t_{eq}) = \frac{ \langle \Phi^{*}(0,t_{eq}) \Phi(z,t_{eq}) \rangle }
{ \sqrt{ \langle |\Phi(0,t_{eq})|^{2} \rangle \langle |\Phi(z,t_{eq})|^{2} 
\rangle }}\;.
\end{equation}
Higher order correlation functions are also routinely obtained by a simple generalisation of the
above formula. In particular, the second and third order correlation functions
evaluated at the same spatial coordinate at equilibrium
are numerically obtained via
\begin{equation}
g^{(2)}(z) = g^{(2)}(z,z,z,z;t_{eq}) = \frac{ \langle |\Phi(z,t_{eq})|^{4} \rangle }
{  \langle |\Phi(z,t_{eq})|^{2} \rangle^{2} }\;,
\end{equation}
and
\begin{equation}
g^{(3)}(z) = g^{(3)}(z,z,z,z,z,z;t_{eq}) = \frac{ \langle |\Phi(z,t_{eq})|^{6} \rangle }
{  \langle |\Phi(z,t_{eq})|^{2} \rangle^{3} }\;,
\end{equation}
respectively.

The above technique is very powerful, as it additionally enables a full non-equilibrium determination
of coherence properties of the system, as demonstrated in \cite{Stoof_Noisy_3}. This is
of direct relevance to recent growth experiments \cite{Low_D_Growth} which will be investigated in
future work.

\subsection{Mean Field Theories}

In our initial analysis, results of the first order correlation function, evaluated from Eq. (5),
 will be compared against
correponding results of a modified finite temperature mean field theory valid in low dimensions 
\cite{Low_D_Theory_2a}. Although good agreement
between these two theories has already been demonstrated \cite{Low_D_Theory_2b}, our aim here is to show 
that the agreement is {\em sufficiently sensitive} to {\em quantitatively} assess the role of density
fluctuations on equilibrium coherence properties.
In particular, under suitable realistic conditions, predictions of the
stochastic approach are found to be in full agreement with the
mean field theory which includes density fluctuations \cite{Low_D_Theory_2a}.
Importantly, however, the same stochastic predictions are shown to
differ substantially from the
limiting case of that mean field theory, in which density fluctuations are ignored
\cite{Low_D_Theory_1}.
This is an important remark, given that all coherence experiments (except \cite{Low_D_Growth})
have so far been analyzed in terms of theories which ignore density fluctuations.

Both mean field theories mentioned above are summarised in Appendix A.
For our present purposes, it is sufficient to quote here the
final expression for the first-order normalized 
correlation function $g^{(1)}(0,z)$, which is given in terms of correlations of the 
phase operator, $\hat{\chi}(z)$  by
 \begin{equation}
g^{(1)}(0,z) =
e^{-{1\over2}\langle\left[\hat{\chi}(z)-\hat{\chi}(0)\right]^2
\rangle}\;.
\label{d11}
\end{equation}   
The exponent of the above expression can be written in 1D as an  appropriate sum over Legendre polynomials,
with an additional prefactor which explicitly depends on the spatial extent of the quasi-condensate
(see Eq. (A1)). The latter parameter is
appropriately defined by a `temperature-dependent
Thomas-Fermi radius', $R_{\rm TF}(T)$, whose size decreases with increasing temperature.

The low temperature theory of Petrov et al. \cite{Low_D_Theory_1} is obtained as a limiting case of
this  theory upon setting the quasicondensate depletion to zero,
as demonstrated in \cite{Low_D_Theory_2d}.
To perform a calculation which is consistent with the `classical' Langevin theory employed here,
we again allow numerically the thermal part to relax to the classical value
$N(\varepsilon) = \left[ \beta (\varepsilon - \mu) \right]^{-1}$. 

Let us now discuss the results of the theories presented above.

\section{First Order Correlation Function at Trap Centre}

\subsection{Temperature Dependence of $g^{(1)}(0,z)$}

The spatial correlation function $g^{(1)}(0,z)$ evaluated at the trap centre at equilibrium by the 
stochastic approach is shown for two different temperatures by the brown (grey) lines in Fig. 1(a)-(b).
Corresponding predictions based on the theory of Andersen et al. \cite{Low_D_Theory_2a}, i.e. including densiy
fluctuations, and the theory of Petrov et al. \cite{Low_D_Theory_1}, i.e. excluding density fluctuations, are shown
respectively by the dashed and solid black lines.
The spatial extent of all plotted curves is restricted to points within the temperature-dependent
Thomas-Fermi (TF) radius $R_{\rm TF}(T)$, which 
characterises the system size at each temperature (see also \cite{Low_D_Theory_2b,Low_D_Theory_2c}).
As expected, the correlation function in the absence of density fluctuations 
is consistently higher than both
other predictions (stochastic theory and mean field theory with density fluctuations), 
with this discrepancy becoming significant at higher temperatures.
Importantly, these latter two theories show consistent behaviour over the entire temperature range,
with only slight differences in the shapes of their predicted correlation functions at intermediate
temperatures, presumably arising because 
the stochastic treatment does not predict gaussian behaviour
close to the origin.

The good agreement between the stochastic approach and the mean field theory with density
fluctuations becomes more evident in Figs. 1(c)-(d), which depict the temperature dependence of the correlation
function suitably characterised by means of two independent determinations. 
Throughout this work, temperature is scaled to the `phase coherence' temperature, 
$T_{\phi} = N (\hbar \omega_{z})^{2}/\mu$ \cite{Low_D_Theory_1}, 
which marks the onset of phase fluctuations and roughly
separates the regions of quasi-condensation and `true' condensation.
Here $N$ denotes the quasi-condensate atom number, $\omega_{z}$ the confining
frequency, and $\mu$ the 1D chemical potential.

The first approach we use here to compare the predictions for this correlation function
over the entire temperature range, based on these three theories,
 is to investigate how the
value of the correlation function halfway to the edge of the quasi-condensate, i.e. at $z=R_{\rm TF}(T)/2$,
varies as a function of temperature. This is shown in Fig. 1(c) against scaled temperature $T/T_{\phi}$.
In the quasi-condensate regime considered here, this quantity
decreases smoothly to zero at a rate which depends critically on 
whether density fluctuations are included (circles, open squares), or not (filled squares),
with both theories including density fluctuations showing very good agreement.

\begin{figure}[t]
\includegraphics[width=8.cm]{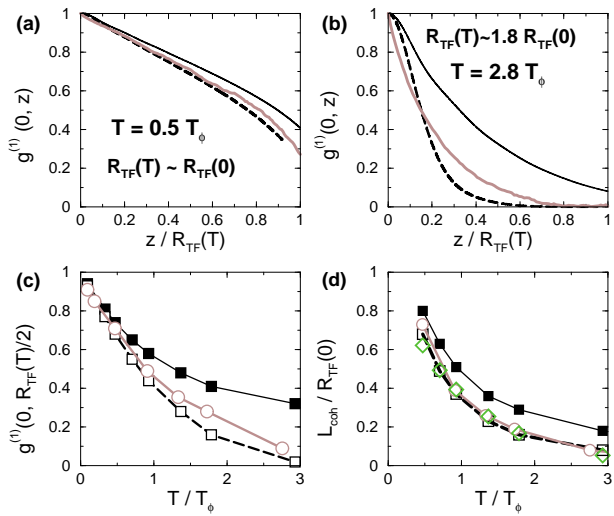}
\caption{(color online)
(a)-(b) Normalized first order spatial correlation function $g^{(1)}(0,z)$ at equilibrium, 
obtained from the stochastic approach (solid brown/grey lines), and from mean-field theory 
with (dashed black) and without (solid black) inclusion of density fluctuations.
Plotted temperatures correspond to $T / T_{\phi} \approx$
(a) $0.5$, and (b) $2.8$, with position $z$ scaled
to the effective temperature-dependent Thomas-Fermi radius $R_{\rm TF}(T)$, the value of which
decreases from (a) to (b).
(The corresponding density profiles are shown in Fig. 6(a)-(b).)
(c) Dependence of $g^{(1)}(0,z=R_{\rm TF}(T)/2)$ on scaled temperature $T/T_\phi$ based on the
stochastic approach (open circles) and modified mean field theory with (open squares) and
without (filled squares) density fluctuations.
(d) Dependence of the scaled coherence length $L_{\rm coh}/R_{\rm TF}(0)$ on scaled temperature $T/T_{\phi}$
for the 3 theories shown in (c). Here 
$L_{\rm coh}$ is defined by $g^{(1)}(0,L_{\rm coh}/R_{\rm TF}(0))=0.5$,
where $R_{\rm TF}(0)$ is the spatial extent of the condensate at $T=0$.
Only points corresponding to temperatures $T>T_{\phi}/2$ are plotted, 
since at lower temperatures the correlation
function remains consistently above the value of $0.5$, indicating full coherence.
Green (grey) diamonds correspond to the curve $L_{\rm coh}/R_{\rm TF}(0) = {\rm exp}(-T/T_{\phi})$.
Plotted lines in (c)-(d) connect data points as a guide to the eye.
}
\end{figure}

An alternative method for characterizing the temperature dependence of the correlation
function is to appropriately extract a coherence length
from it, scale this to the system size, and study its variation as a function
of scaled temperature $T/T_{\phi}$.
A graph of this form is often compiled by experimentalists \cite{Low_D_ENS_EPJD}.
Here we choose to define the coherence length, $L_{\rm coh}$, 
as the value of $z$ at which the correlation function
decays to half its original value, i.e. $g^{(1)}(0,L_{\rm coh})=0.5$.
We shall deal with a more suitable definition of the coherence length which also
accounts for the changing shape of the correlation function in Sec. III C.
In our numerical results, $L_{\rm coh}$ is scaled to the zero-temperature system size $R_{\rm TF}(0)$, and
its dependence on  scaled
temperature $T/T_{\phi}$ is shown in Fig. 1(d).
Importantly, we find that all curves which include density fluctuations lie on a universal curve
$L_{\rm coh}/R_{\rm TF}(0)={\rm exp}(-T/T_{\phi})$, whereas curves without density fluctuations (filled squares)
lie considerably higher, indicating an overestimate of the actual equilibrium coherence length.
As argued elsewhere \cite{Low_D_Theory_2d}, the shift between curves which include the coupling of
phase and density fluctuations to those that do not, depends critically on the ratio of the phase coherence
temperature $T_{\phi}$, to the 1D `transition temperature' $T_{c}$.
Importantly, however, we note that the trend of the observed decrease of the coherence length due to density fluctuations
is consistent with recent experimental findings \cite{Low_D_ENS_EPJD}.

All results presented in this paper are plotted in terms of suitably scaled parameters
(lengths, temperatures), such that any dependence on the particular system details is removed.
The underlying simulations were performed at constant chemical potential $\mu=30 \hbar \omega_{z}$
for a 1D gas of $^{23}$Na atoms, of scattering length $a_{\rm 3D}=2.75$nm, under
 longitudinal confinement of 
frequency $\omega_{z}=2\pi \times 3.5$ Hz.
The 1D interaction strength, denoted by $g$ in Eq. (1), is given by $g=4 \pi \hbar^{2} \kappa /m$, 
where $\kappa=(m/2 \pi \hbar) \omega_{\perp} a_{\rm 3D}$ is the 1D 
`coupling constant' (or `inverse scattering length') \cite{Olshanii}. 
In our 1D simulations the value of $\kappa$ is fixed by our choice of
transverse confinement, taken here to be harmonic with frequency
$\omega_{\perp}=2 \pi \times 120$Hz.
%
Performing the simulation at fixed chemical potential leads to a slight variation in the 
atom numbers in the trap, in the range $18,500-24,000$, for the considered temperature range,
$10$nK $< T < 400$ nK.
An analogous variation would also be expected in experimental realisations at different temperatures
and approximately fixed atom numbers.
Accordingly, in the considered temperature range, the phase coherence temperature $T_{\phi}$ 
increases approximately linearly with increasing atom number  
(or, equivalently here, temperature $T$), from its initial value $T_{\phi} \sim 100$ nK to approximately
$T_{\phi} \sim 125$ nK.

\subsection{Symmetrically Evaluated Correlation Function $g^{(1)}(-z/2,z/2)$}

A number of current experiments are better suited to 
determining the correlation
function symmetrically about a point, as done in the 
Hannover \cite{Low_D_Exp_3,Low_D_Hannover_PRA} and Orsay experiments \cite{Low_D_Exp_4}.
Our next task is therefore to investigate the extent to which the predictions discussed above depend on the precise method
by which the correlation function is obtained.
We will do this by comparing corresponding predictions at the trap center, 
based on the stochastic theory.

\begin{figure}[b]
\includegraphics[width=8.cm]{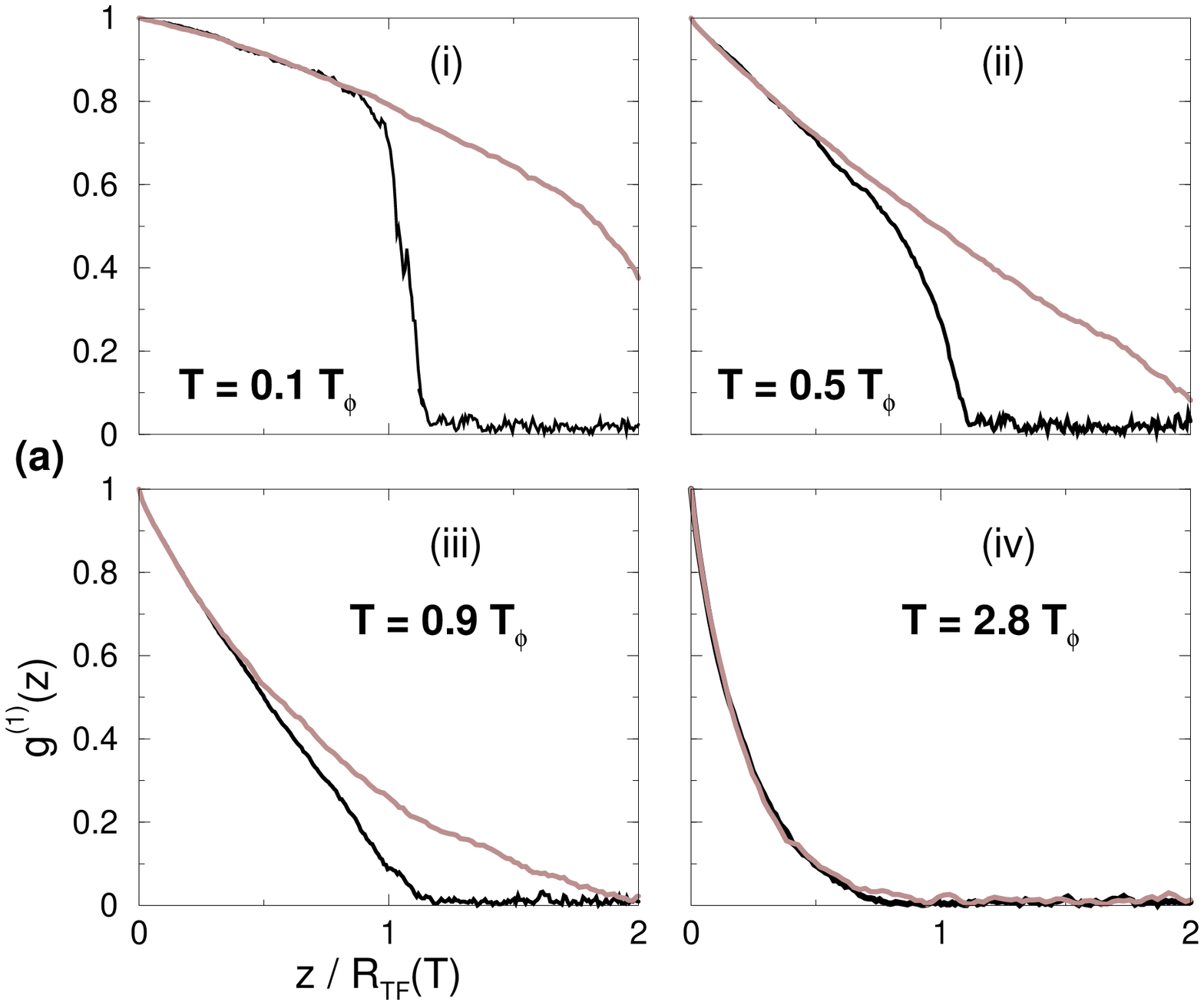}
\includegraphics[width=8.cm]{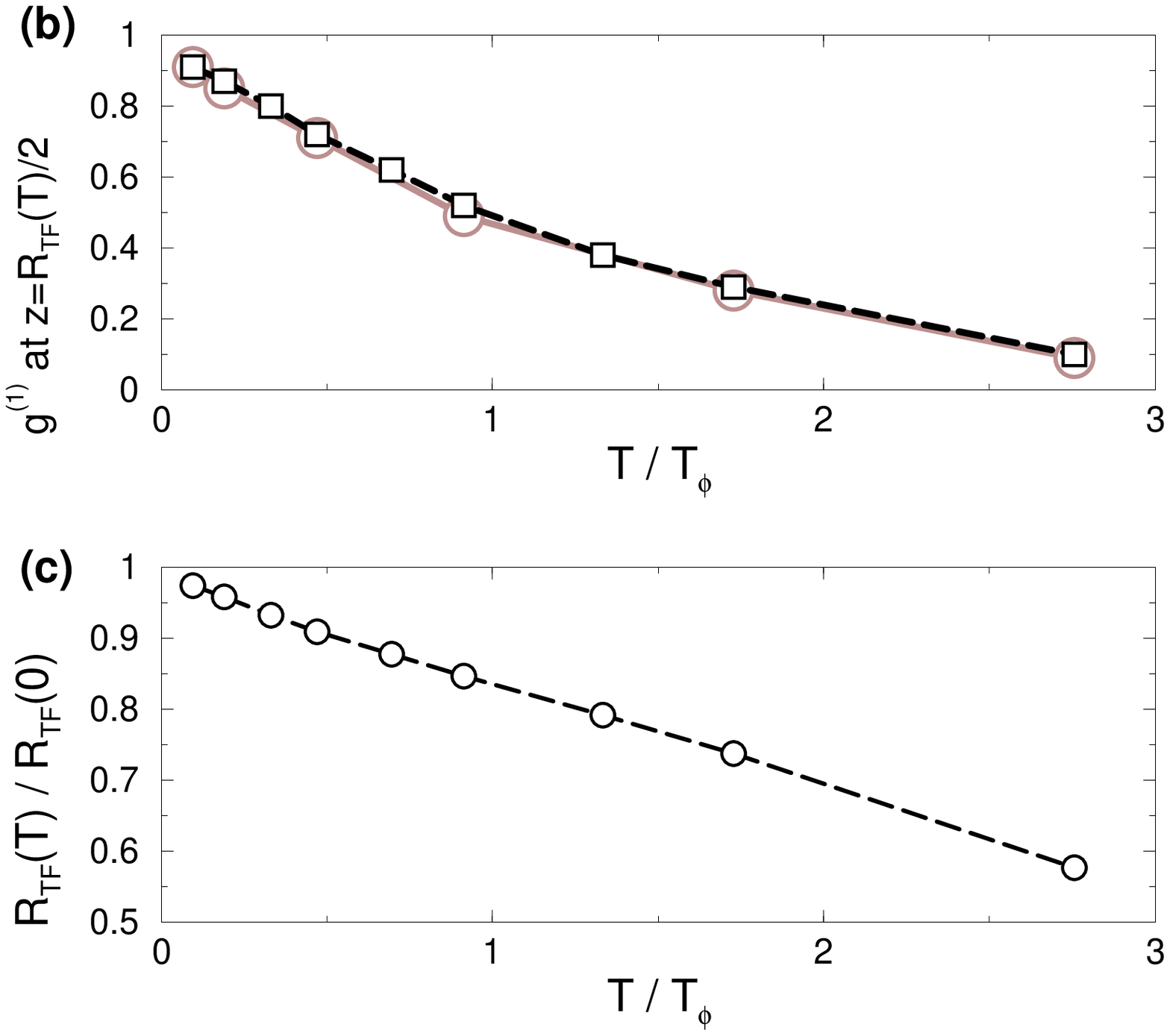}
\caption{(color online) 
(a) (i)-(iv) Spatial dependence of the normalised first order correlation function,
calculated via the stochastic theory, under different autocorrelation measurements.
Black lines denote the function  $g^{(1)}(0,z)$ of Eq. (5), whereas brown (grey) lines show the
symmetrically evaluated  $g^{(1)}(-z/2,z/2)$ of Eq. (9).
Plotted graphs correspond to temperatures  $T/T_{\phi} \approx$
(i) 0.1, (ii) 0.5, (iii) 0.9, and (iv) 2.8, with
$z$ scaled throughout to the temperature-dependent system size $R_{\rm TF}(T)$.
(b) Value of the normalised first-order correlation function 
evaluated at $z=R_{\rm TF}(T)/2$ from the trap centre, 
as a function of scaled temperature $T/T_{\phi}$, 
obtained from $g^{(1)}(0,z)$ (brown/grey circles),
and $g^{(1)}(-z/2,z/2)$ (black squares).
(c) Dependence of $R_{\rm TF}(T)$ on scaled temperature $T/T_{\phi}$.
Plotted lines in (b)-(c) are a guide to the eye.
}
\end{figure}

The `symmetric' correlation function at the trap centre is thus
obtained at equilibrium via the following numerical autocorrelation
\begin{equation}
g^{(1)}(-z/2,z/2;t_{eq}) = \frac{ \langle \Phi^{*}(-z/2,t_{eq}) \Phi(z/2,t_{eq}) \rangle }
{ \sqrt{ \langle |\Phi(-z/2,t_{eq}) |^{2} \rangle \langle |\Phi(z/2,t_{eq}) |^{2} \rangle }}
\end{equation}
which should be compared and contrasted to Eq. (5).

Correlation functions computed symmetrically via Eq. (9) are shown by the brown (grey) lines
in Fig. 2(a) for four different values of $T/T_{\phi}$.
These are compared to corresponding
correlation functions $g^{(1)}(0,z)$ of Eq. (5) (black).
We find that, for a given temperature which fixes the density profile, 
these two correlation functions 
are indistinguishable in a broad region close to the trap centre.
Some differences do, however, arise towards the edge of the quasi-condensate, and beyond, 
for certain temperatures just below
the phase coherence temperature $T_{\phi}$.
A slower decrease of the symmetrically computed correlation function
$g^{(1)}(-z/2,z/2)$ is anticipated in this region, since the different
 methods of evaluation used imply that the considered 
correlation functions will decay to zero at different points,
respectively at $z \approx R_{\rm TF}(T)$ for $g^{(1)}(0,z)$ and $z \approx 2R_{\rm TF}(T)$ 
for $g^{(1)}(-z/2,z/2)$.

For sufficiently low temperatures,  as in Fig, 2(a)(i),
the agreement between the two different
determinations of the correlation function is accurate almost up to the edges of the
quasi-condensate (i.e., $z=R_{\rm TF}(T)$),
and the same holds deeply in the quasi-condensate regime shown in Fig. 2(a)(iv), for which the 
two independently determined correlation
functions are practically indistinguishable.
In fact, this equivalence remains nearly perfect over the entire temperature range, 
for spatial coordinates $z \le R_{\rm TF}(T)/2$.
This is evident in Fig. 2(b) which compares the temperature dependence of the value of the correlation
function at $z=R_{\rm TF}(T)/2$, based on the two different determinations mentioned above.
The reader is reminded that a similar determination of the coherence length was used in
Fig. 1(c) to demonstrate the importance of density fluctuations.
For completeness, Fig. 2(c) plots the effective system size, $R_{\rm TF}(T)$, versus 
scaled temperature $T/T_{\phi}$, demonstrating an approximately
linear decrease with increasing temperature.

One might also consider performing the alternative analysis of Fig. 1(d), whereby
the coherence length is defined by the spatial coordinate $z$, at which the first
order correlation function $g^{(1)}$ decays to the value of $0.5$, with this value scaled to $R_{\rm TF}(0)$.
In this case,  at low temperatures
$T \le T_{\phi}$, the two correlation functions
are sensitive to their precise evaluation method, and the agreement between them would not be as good.

\subsection{Exponential vs. Gaussian Profiles}

\begin{figure}[t]
\includegraphics[width=8.cm]{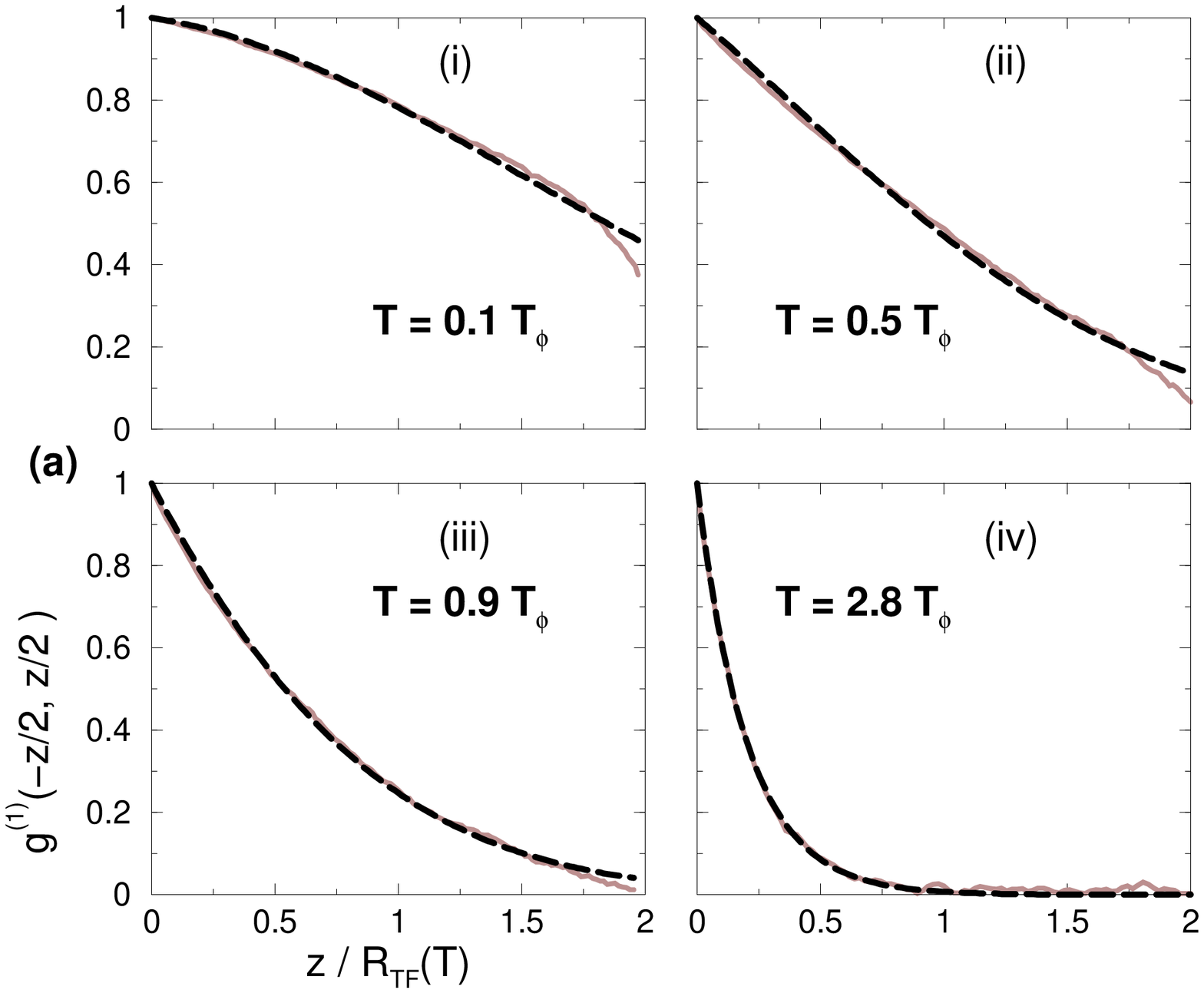}
\includegraphics[width=8.cm]{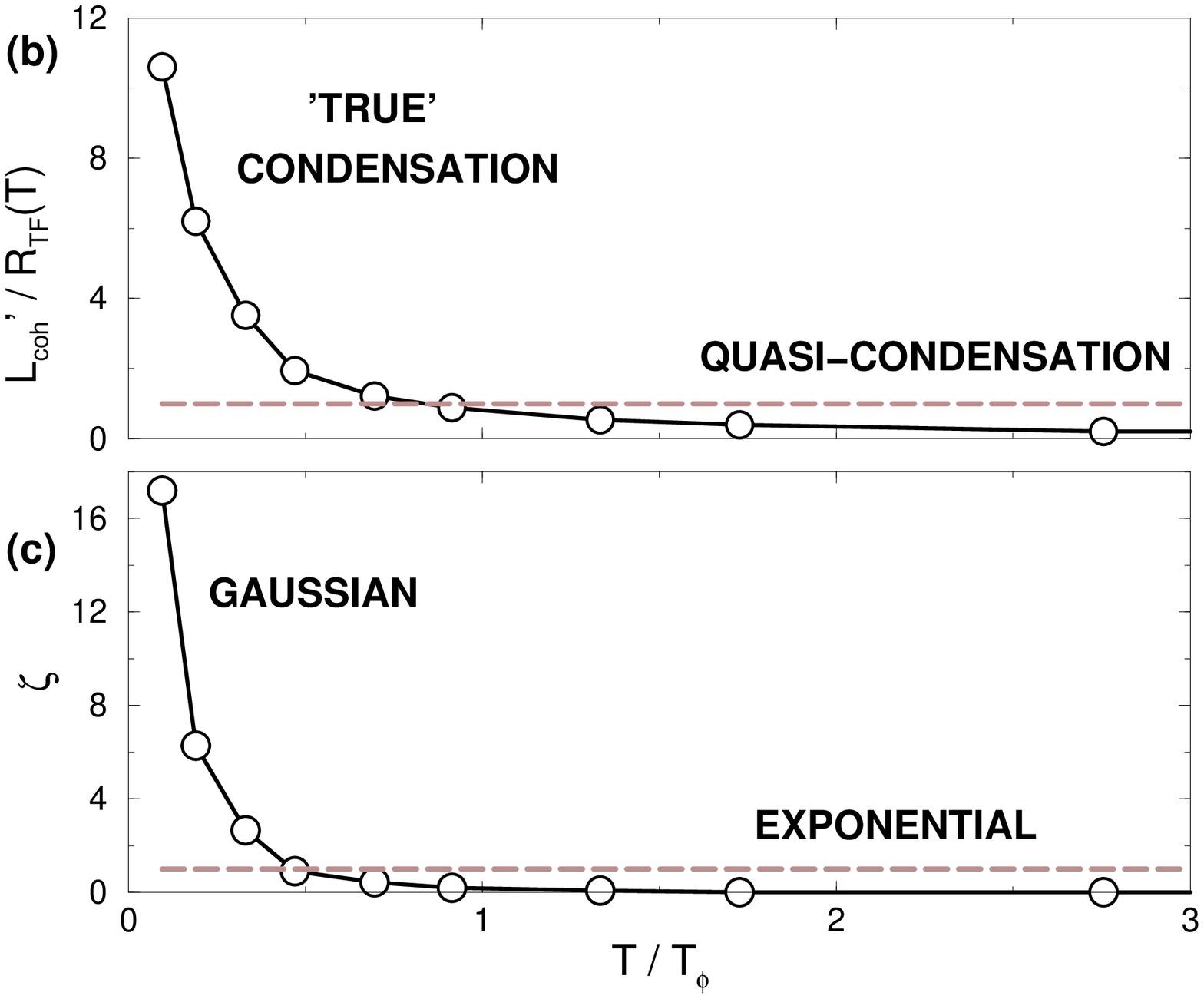}
\caption{(color online) 
(a) (i)-(iv) Symmetrically computed spatial correlation function
$g^{(1)}(-z/2,z/2)$, where $z$ is scaled to the temperature-dependent
Thomas-Fermi radius $R_{\rm TF}(T)$ (solid brown/grey), versus numerical fit by the function $f(z)$ of
Eq. (10) (dashed black), for the graphs of Fig. 2(a).
(b) Temperature dependence of the coherence length $L_{\rm coh}'$ (black circles) obtained by above fit,
with $L_{\rm coh}'$ scaled to $R_{\rm TF}(T)$, and temperature to $T_{\phi}$.
The dashed brown (grey) line highlights a coherence equal to the system size, i.e. $L_{\rm coh}'=R_{\rm TF}(T)$.
(c) Corresponding dependence of the `crossover' parameter $\zeta$ (black circles)
determining the relative importance of exponential
and gaussian contributions to the spatial correlation.
The dashed brown (grey) line separates regions of predominantly gaussian behaviour ($\zeta \gg 1$),
to those of exponential behaviour. Plotted solid lines in (b)-(c) are a guide to the eye. 
}
\end{figure}

The presented analysis thus far has focused only on the `global' loss of coherence with increasing
temperature, 
without paying attention to the precise shape of the correlation function.
The latter is known to vary from exponentional to gaussian 
with decreasing temperature, indicating the crossover from quasi-condensation 
to `true' condensation \cite{Low_D_Theory_3d}.
In this section we study the temperature crossover in the shape of this function in more detail,
by means of our
previously computed symmetric correlation function $g^{(1)}(-z/2,z/2)$.
Since $g^{(1)}$ is known to be gaussian at $T=0$ and exponential at temperatures $T>T_{\phi}$,
we fit it, in the relevant region $z \le 2 R_{\rm TF}(T)$,  
by a function $f(z)$ which provides a smooth crossover between the two shapes, of the form
\begin{equation}
f(z) = e^{-\left[ (z/L_{\rm coh}') + \zeta (z/L_{\rm coh}')^{2} \right] }\;.
\end{equation}
The parameter $\zeta$ appearing above is a measure 
of how `gaussian' or `exponential' a particular  profile is,
while the chosen fit provides a unique definition of the coherence length, $L_{\rm coh}'$, 
which decreases monotonically with increasing temperature, irrespective of the change in the profile's
shape.

The correlation functions $g^{(1)}(-z/2,z/2)$ for the four different temperatures discussed earlier are shown
by the brown (grey) lines in Fig. 3(a); the corresponding fits based on Eq. (10)
are plotted on the same figure by dashed black lines.
Such fits enable us to study the dependence of the `crossover parameter', $\zeta$, and the
coherence length, $L_{\rm coh}'$, on temperature. In particular,
the scaled coherence length $L_{\rm coh}'/R_{\rm TF}(T)$ and the `crossover' parameter
$\zeta$ are plotted against scaled temperature $T/T_{\phi}$ in Fig. 3(b)-(c).
Note that a value of $\zeta = 0$ corresponds to a purely exponential correlation function, with the gaussian
limit reached for $\zeta \gg 1$.

At low temperatures, $T < T_{\phi}$, we find that the coherence length $L_{\rm coh}'$
is comparable to, or larger than the quasi-condensate spatial extent, $R_{\rm TF}(T)$,
for that particular temperature, signalling the appearance of `true' condensation.
In this regime, the gaussian contribution has a larger weight, than the
exponential term. Moreover, we find that 
the ratio in their relative contributions decreases with increasing temperature, 
with $\zeta \approx 1$ at the `crossover region'. 
Finally, as the temperature increases further, the system enters
the quasi-condensate regime and the
correlation function essentially acquires an exponential profile
 for temperatures $T>T_{\phi}$, consistent with experimental observations.


\section{Higher Order (Density-Density) Correlation Functions}

In this section we consider higher order correlation functions, which contain information about
density-density correlations.
These are considered at fixed points $z$ and at equilibrium, with the subsequent analysis focusing on
the correlation functions
$g^{(2)}(z)=g^{(2)}(z,z,z,z;t_{eq})$, and $g^{(3)}(z)=g^{(3)}(z,z,z,z,z,z;t_{eq})$ of Eqs. (6)-(7)
as a function
of $z$ for various temperatures.

\subsection{Dependence on Temperature}

The behaviour of $g^{(2)}(z)$ and $g^{(3)}(z)$ against $z$ is shown for various temperatures 
in Figs. 4(a) and (b), with position scaled to the zero-temperature Thomas-Fermi radius, $R_{\rm TF}(0)$.
As expected, these functions have a lower bound of $1$, occuring deep within the quasi-condensate region
($z \ll R_{\rm TF}(0)$), and at sufficiently low temperatures, demonstrating full coherence.
The corresponding upper bounds for these functions are $g^{(2)}(0)=2!=2$, and   $g^{(3)}(0)=3!=6$,
consistent with the observed values in 3D BECs \cite{g2a,g2b,g2c,g3}.
These values correspond to complete absence of coherence, 
and such incoherent thermal atoms are typically located
outside the quasi-condensate region. 

In particular, as $T \rightarrow 0$, both functions $g^{(2)}(z)$ and $g^{(3)}(z)$
tend towards step-like functions $g^{(n)}(z)=1+(n!-1)\Theta(z)$ centered around $R_{\rm TF}(0)$, 
with $\Theta(z) =$ $0$ for $z \le R_{\rm TF}(0)$ and $1$ for $z>R_{\rm TF}(0)$. As the temperature increases, we observe the 
following features, which are consistent with the results of \cite{Davis_Class_Trap}: 
Firstly, the central values $g^{(2)}(0)$ and $g^{(3)}(0)$ increase due to the increasing thermal component
located at the trap centre; their dependence on temperature is shown in Fig. 4(c).
Furthermore, the crossover between the quasi-condensate-dominated region $z<R_{\rm TF}(T)$, and the purely
thermal region $z>R_{\rm TF}(T)$ becomes smoother,
due to the increased presence of thermal atoms over the entire trap extent.
Finally, the location of this crossover is shifted to smaller values of $z/R_{\rm TF}(0)$,
consistent with the decrease in the spatial extent $R_{\rm TF}(T)< R_{\rm TF}(0)$ of the quasi-condensate
with increasing temperature.

\begin{figure}[t]
\includegraphics[width=8.cm]{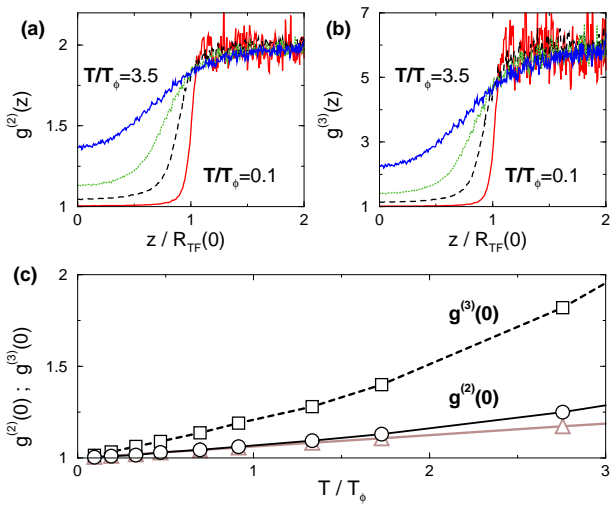}
\caption{(color online)
(a)-(b) Spatial dependence of the
(a) second, $g^{(2)}(z)$, and (b) third order, $g^{(3)}(z)$, equilibrium spatial correlation functions at the same point $z$,
for various temperatures; from bottom to top $T/T_{\phi} \approx$: (i) $0.1$, (ii) $0.7$, (iii) $1.7$,
and (iv) $3.5$. 
Distances are scaled to the zero temperature system size $R_{\rm TF}(0)$.
(c) Corresponding values of $g^{(2)}(0)=g^{(2)}(0,0,0,0;t_{eq})$ (black circles) 
and $g^{(3)}(0)=g^{(3)}(0,0,0,0,0,0;t_{eq})$ (black squares)
at trap centre as a function of temperature.
Brown (grey) triangles indicate corresponding results for $g^{(2)}(0)$ based on Eq. (11).
Plotted lines in (c) are a guide to the eye.
}
\end{figure}

Our findings for the temperature dependence of $g^{(2)}(0)$ in Fig. 4(c) are further compared to the analytical result of Kheruntsyan et al. \cite{Kheruntsyan_g2_1,Kheruntsyan_g2_2}.
In their work, the interaction strength is parametrised in terms of the parameter $\gamma = mg/\hbar^{2}n$,
where $m$ is the atomic mass, $g$ the one-dimensional coupling constant, and $n$ the density of the gas.
In the regime $\gamma \ll 1$ considered here, they find the following temperature dependence
(see Eq. (5.10) in Ref. \cite{Kheruntsyan_g2_2})
\begin{equation}
g^{(2)}(0) = 1 + \frac{4 \sqrt{2}}{3} \left( \frac{ T}{T_{d}} \right)
\end{equation}
where $T_{d}$ is the degeneracy temperature, defined by $T_{d}=N \hbar \omega_{z}$, $N$ is the number
of quasi-condensate atoms and $\omega_{z}$ is the confining frequency.
We apply this analytical formula to our numerical results by extracting the relevant
parameter $(T/T_{d})$ from our stochastic simulations.
The corresponding temperature dependence of $g^{(2)}(0)$ predicted by Eq. (11) is thus plotted by
 the brown (grey) triangles  in Fig. 4(c). Thus there is good agreement between the stochastic and analytical theories
 in the temperature range considered here. However, the stochastic theory
predicts an additional 
slight {\em reduction} in coherence, as manifested by the {\em higher} value of $g^{(2)}(0)$,
with the extent of this reduction increasing with increasing temperature.

\subsection{Dependence on Interaction Strength}

Next, we investigate the dependence of the three-lowest order correlation functions on the effective 1D 
interaction strength at fixed temperature, with our analysis again restricted 
to the weakly-interacting regime $\gamma \ll 1$. 
As mentioned earlier, the 1D interaction strength can be parametrised in terms of the 
1D coupling constant  $\kappa = (m/2 \pi \hbar) a_{\rm 3D} \omega_{\perp}$.
This parameter can be changed either by tuning the 3D scattering length 
via Feshbach resonances \cite{Feshbach_1,Feshbach_2}, or by modifying the transverse
confinement, while keeping all other parameters fixed.
Since an increase in the interaction strength causes a reduction of coherence, as demonstrated explicitly
below, we have chosen to discuss here a rather low temperature example, with $T=0.5T_{\phi}$, for which
there is appreciable coherence in the system.

The dependence of the three lowest order correlation functions on coupling constant $\kappa$ is
shown in Fig. 5, with the coupling constant used thus far in our discussion, and in Figs. 1-4,
henceforth denoted by $\kappa_{0}$.
Since temperature is kept fixed for all curves shown in this section,
we have chosen here to scale distances to the quasi-condensate spatial extent at the given
temperature $T=0.5T_{\phi}$, based on the coupling constant $\kappa_{0}$.
For this particular coupling constant, 
the corresponding first order correlation functions were shown in Figs. 2(a)(ii) and 3(a)(ii)
(note the different plot ranges),
whereas higher order correlation functions were plotted by the dashed black lines in Figs. 4(a)-(b).

The change in the correlation function $g^{(1)}(-z/2,z/2)$
as a function of position for 
different values of the 1D coupling constant
is shown in Fig. 5(a). As evident,
an increase in $\kappa$, from its
previously considered value $\kappa_{0}$,
 leads to a decrease in the coherence of the system. This is clearly portrayed in Fig. 5(b) showing the value of the
correlation function at $z=R_{\rm TF}(T)/2$ as a function of $\kappa$.
This curve further demonstrates that 
full coherence is asymptotically
approached in the limit of negligible interactions, i.e. as $\kappa \rightarrow 0$, as anticipated.

Figs. 5(c)-(d) show corresponding dependence of $g^{(2)}(z)$ and $g^{(3)}(z)$ on $z$ for different interaction
strengths, with top curves corresponding to a ten-fold increase in $\kappa$.
Our analysis reveals that an increase of the interaction strength leads to an increase in the central
values of the correlation function  $g^{(2)}(0)$ and $g^{(3)}(0)$, and a more
gradual crossover to incoherent behaviour outside the quasi-condensate region.
Thus, qualitatively,
an increase in the interaction strength produces the same effect on the
system coherence as an increase in the temperature of the system.

\begin{figure}[t]
\includegraphics[width=8.cm]{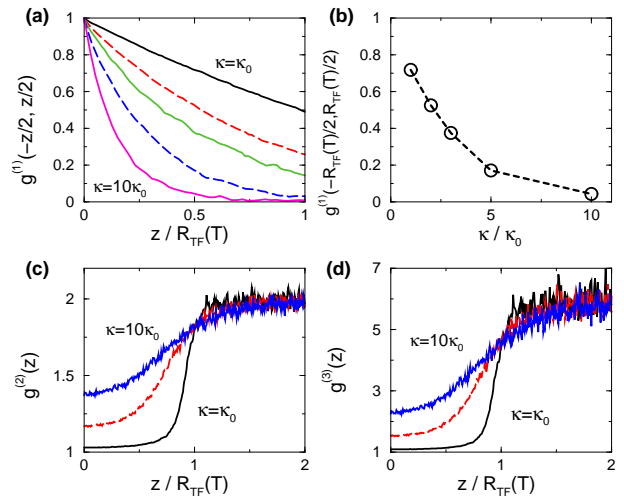}
\caption{(color online) 
Dependence of the three lowest order correlation functions on the effective 1D 
coupling constant $\kappa$ at equilibrium for a fixed low temperature
$T \approx 0.5 T_{\phi}$:
(a) Dependence of $g^{(1)}(-z/2,z/2)$ on $\kappa$ 
as a function of position, for
(from top to bottom) $\kappa/\kappa_{0}=1,2,3,5$ and $10$, where
$\kappa_{0}$ is the value of the effective 1D coupling constant used in Figs. 1-4.
(b) Dependence of $g^{(1)}(-z/2,z/2)$  at $z=R_{\rm TF}(T)/2$ on $\kappa/\kappa_{0}$, with the dashed line
simply connecting the data points.
(c)-(d) Second and third order spatial correlation functions at the same point and at equilibrium
as a function of position, for an interaction strength
(from bottom to top) $\kappa/\kappa_{0} =1,5,10$.
Here, the value of the 1D coupling constant
$\kappa$ has been changed by modifying the transverse confinement
from $\omega_{\perp}=2 \pi \times 120$Hz, to the values $\omega_{\perp}'/2 \pi= 240, 360, 600$, and $1200$ Hz.
In this figure, all positions have been scaled to the spatial extent of the system at $T=0.5T_{\phi}$
for a coupling constant $\kappa=\kappa_{0}$. 
}
\end{figure}

\section{Direct Determination of Quasi-condensate Profiles}

The calculation of density-density correlations discussed previously
enables the precise determination of the quasi-condensate density, to which we now turn our attention.

\subsection{Methodology}

To be able to identify the quasi-condensate part in our density profiles, we use the fact that
the Langevin field $\Phi(z,t)$ essentially contains the physics of the 
Bose field operator $\hat{\Psi}(z,t)$. Such a correspondence has already been used in numerically
evaluating the three lowest order correlation functions $g^{(1)}(\cdots)$, $g^{(2)}(\cdots)$,
and $g^{(3)}(\cdots)$ from Eqs. (5)-(7) in Secs. III and IV.
At a given time, the total atomic density is obtained via
\begin{equation}
n(z)=\langle \hat{\Psi}^{\dag}(z)  \hat{\Psi}(z) \rangle 
\rightarrow \langle \left| \Phi(z) \right|^{2} \rangle \;,
\end{equation}
where the latter averaging is performed over different realizations of the noise.

In order to separate out quasi-condensate and thermal contributions, we
impose the usual `decomposition' of the Bose field operator into a `mean field part'
$\psi_{0}(z)$ and a `fluctuating part' $\hat{\delta}(z)$, i.e.,
\begin{equation}
\hat{\Psi}(z) = \langle \hat{\Psi}(z) \rangle + \hat{\delta}(z) = \psi_{0}(z) + \hat{\delta}(z)\;.
\end{equation}
This directly enables us to split the atomic profile of Eq. (12) 
into a `quasi-condensate part', $n_{\rm QC}(z)$, and a `thermal part',
$n_{\rm T}(z)$, via 
\begin{equation}
n(z)=n_{\rm QC}(z)+n_{\rm T}(z)=\left| \psi_{0}(z) \right|^{2}
+ \langle \hat{\delta}^{\dag}(z) \hat{\delta}(z) \rangle\;.
\end{equation}

We now seek to re-write these two components in terms of expressions involving same-time averages of multiple
Bose field operators, since the latter quantities 
can be routinely evaluated within our numerical scheme.
In particular, we note that the average over four Bose field operators at the same spatial coordinate (and same time)
can be written as
\begin{eqnarray}
& & \hspace{-1.5cm} \langle \hat{\Psi}^{\dag}(z) \hat{\Psi}^{\dag}(z)  \hat{\Psi}(z) \hat{\Psi}(z) \rangle
= \left| \psi_{0}(z) \right|^{4} \nonumber \\
& & + 4 \left| \psi_{0}(z) \right|^{2} \langle \hat{\delta}^{\dag}(z) \hat{\delta}(z) \rangle 
+ 2 \langle \hat{\delta}^{\dag}(z) \hat{\delta}(z) \rangle^{2} 
\end{eqnarray}
where  we have used Wick's theorem in the form 
\begin{eqnarray}
\langle \hat{\delta}^{\dag}(z)  \hat{\delta}^{\dag}(z) \hat{\delta}(z) \hat{\delta}(z) \rangle
= 2 \langle \hat{\delta}^{\dag}(z) \hat{\delta}(z) \rangle^{2}\;.
\end{eqnarray}
At the same time, we make use of the following correspondence
\begin{eqnarray}
\langle \hat{\Psi}^{\dag}(z) \hat{\Psi}^{\dag}(z)  \hat{\Psi}(z) \hat{\Psi}(z) \rangle 
& \rightarrow & \langle \Phi^{*}(z) \Phi^{*}(z) \Phi(z) \Phi(z) \rangle \nonumber \\
& = & \langle \left| \Phi(z) \right|^{4} \rangle
\end{eqnarray}
So, from Eqs. (12)-(17) it is easy to see that, in the region $z \le R_{\rm TF}(T)$ where a quasi-condensate exists,
and therefore the above splitting into mean field and fluctuations
is meaningful, the following relation holds
\begin{equation}
\left| \psi_{0}(z) \right|^{4} = 2 \langle \hat{\Psi}^{\dag}(z) \hat{\Psi}(z) \rangle^{2}
- \langle \hat{\Psi}^{\dag}(z) \hat{\Psi}^{\dag}(z)  \hat{\Psi}(z) \hat{\Psi}(z) \rangle\;.
\end{equation}

The quasi-condensate density $n_{\rm QC}(z) = \left| \psi_{0}(z) \right|^{2}$ can hence be accurately determined 
in the range $z \le R_{\rm TF}(T)$ from our numerical scheme via
\begin{eqnarray}
n_{\rm QC}(z) & = & \sqrt{ 2 \langle \hat{\Psi}^{\dag}(z) \hat{\Psi}(z) \rangle^{2}
- \langle \hat{\Psi}^{\dag}(z) \hat{\Psi}^{\dag}(z)  \hat{\Psi}(z) \hat{\Psi}(z) \rangle} \nonumber \\
& \rightarrow &  \sqrt{ 2 \langle |\Phi(z)|^{2} \rangle^{2} - \langle |\Phi(z)|^{4} \rangle }\;.
\end{eqnarray}
Correspondingly, the profile of the thermal cloud, 
$n_{\rm T}(z)=\langle \hat{\delta}^{\dag}(z) \hat{\delta}(z) \rangle$ is directly obtained via
\begin{equation}
n_{\rm T}(z) = \langle |\Phi(z)|^{2} \rangle -n_{\rm QC}(z)\;.
\end{equation}
This approach has already been used (but not commented upon) in parallel work discussing quasi-condensate 
growth on an atom chip \cite{Dimple_Growth_PRA}.

Note that an alternative approach to determine quasi-condensate density profiles has been recently 
discussed in
\cite{Davis_Class_Trap}, based on the projected Gross-Pitaevskii scheme \cite{Davis_Class_PRL}.

\begin{figure}[b]
\includegraphics[width=8.cm]{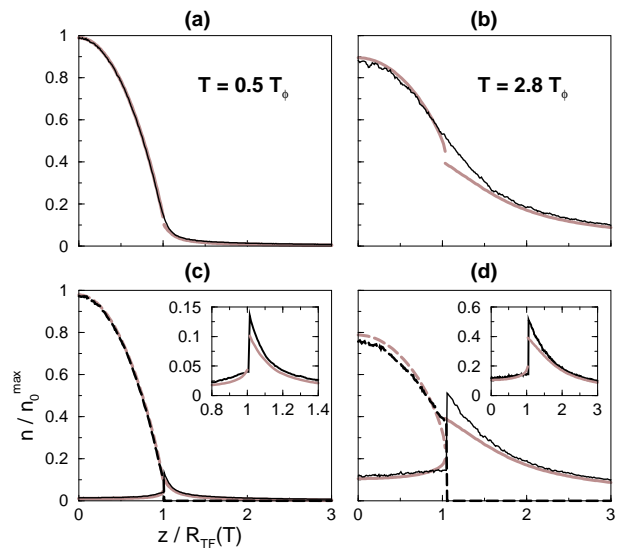}
\caption{
(a)-(b) Total density profiles, scaled to the $T=0$ peak central density, $n_{0}^{\rm max}$,
 as obtained from the
stochastic approach (black), and mean field theory with density fluctuations (brown/grey) 
corresponding to the correlation functions plotted Fig. 1(a)-(b).
(c)-(d) Corresponding quasi-condensate (dashed) versus thermal cloud profiles (solid).
Insets: Thermal cloud profiles around their maxima, $z=R_{\rm TF}(T)$.
All graphs are scaled to the temperature-dependent Thomas-Fermi radius $R_{\rm TF}(T)$
for the particular temperature, and are based
on the coupling constant $\kappa_{0}$ of Figs. 1-4.
}
\end{figure}

\subsection{Results}


To apply the above method, we start by plotting the
total density profiles at two different temperatures, above and below $T_{\phi}$ in Fig. 6(a)-(b).
These figures depict
the good agreement between the stochastic and mean field approaches commented upon in earlier work
\cite{Low_D_Theory_2b}.
In this paper such an analogy is extended much further, by directly comparing quasi-condensate 
and thermal profiles predicted by these two theories, as shown in Fig. 6(c)-(d).
The quasi-condensate density is evaluated in the stochastic theory
by Eq. (19) within the range $0 \le z \le R_{\rm TF}(T)$, and is zero elsewhere.

In order to make a meaningful comparison between these distinct approaches, 
at any given temperature, the same
temperature-dependent Thomas-Fermi radius is used for both theories, with $R_{\rm TF}(T)$
evaluated from the mean field theory.
Dashed lines in Figs. 6(c)-(d) 
depict the quasi-condensate, as evaluated from the stochastic (black) and the mean field
theory (brown/grey), whereas solid lines show corresponding thermal cloud distributions.
The latter are highlighted in the insets
around their respective maxima, located at the quasi-condensate edges $z=R_{\rm TF}(T)$.
The agreement between the predicted profiles is excellent, and remains so over the entire
temperature range considered.

The approach presented here further enables the direct determination of the equilibrium
quasi-condensate fraction at any temperature. In particular, we note that the quasi-condensate fraction is
found to decrease approximately linearly with increasing temperature.

\subsection{Application to Experiments}

Experiments with quasi-condensates typically rely on bimodal density fits to determine important
parameters such as the quasi-condensate spatial extent and quasi-condensate fraction
at a given temperature.
The procedure described above provides a direct alternative way of obtaining an unequivocal, theory-independent
determination of quasi-condensate and thermal density profiles in quasi-1D experiments.

In order to perform such an analysis experimentally, one must simultaneously obtain both
the total atomic density profiles $n(z)=\langle | \Phi(z)|^{2} \rangle$, as done routinely, 
as well as the density-density correlations $\langle n^{2}(z) \rangle = \langle |\Phi(z)|^{4} \rangle$
at various points across the entire trap. This experiment should ideally be performed in situ,
which is within current experimental reach \cite{Low_D_ENS_Density_Fluctuations}, 
in order to avoid the
coupling of density and phase fluctuations occuring in the usual time-of-flight expansion stage \cite{Low_D_Exp_1,Low_D_Exp_4}.

\section{Conclusions}

In conclusion, we carried out a systematic analysis of
the dependence of the three lowest order correlation functions 
of a one-dimensional
ultracold atomic gas on temperature and effective interaction strength,
by means of a stochastic Langevin approach in the `classical' approximation.
Our discussion was limited to the weakly-interacting regime ($\gamma \ll 1$),
which features a smooth temperature crossover from an incoherent gas, to a quasi-condensate,
and, finally, at sufficiently low temperatures, to a `true' condensate.
This work complements earlier work in this area, which studied the interplay between density
and phase fluctuations.

Results for the off-diagonal normalized first order correlation function were shown to
be practically indistinguishable from those of a mean field theory which is valid in low dimensions
and explicitly includes density fluctuations.
This should be contrasted to theories which a priori ignore density fluctuations, with such theories
shown to predict a
larger amount of coherence than is physically present. 
Two different approximations for the evaluation of this correlation function were
used, corresponding to different interference experiments that could be performed, 
and their profiles were shown to be very similar within the appropriate spatial range.
We further investigated the temperature crossover in the shape of the correlation function, which evolves
from gaussian to exponential with increasing temperature, and used this to obtain a monotonically
varying coherence length.

Moreover, second and third order correlation functions at the same point were investigated as a
function of temperature and interaction strength. The observed decrease in coherence when either
of these parameters is increased was shown to be consistent with other results in the literature.
Importantly, consideration of density-density correlations was shown to lead to an accurate determination of the
quasi-condensate and thermal density profiles.
This approach could be used in experiments, to provide a more accurate determination of
quasi-condensate profiles and fractions, offering an alternative to 
the bimodal fitting schemes currently used.

In addition to providing valuable information for equilibrium profiles,
 to which the present analysis was restricted, the stochastic approach is well-suited 
for discussing growth of coherence in quasi-one-dimensional Bose gases, an issue that will be
addressed in subsequent work.

\acknowledgments
It is a pleasure to acknowledge numerous discussions with Henk Stoof which have significantly
contributed to the present work.

\appendix \section{Low-dimensional Mean Field Theories}

This Appendix summarizes the modified finite temperature mean field theory of
Andersen, Al Khawaja and Stoof \cite{Low_D_Theory_2a}.
This theory extends alternative approaches suited to low-dimensional gases 
\cite{Low_D_Theory_1,Low_D_Theory_3a} by providing an ab initio self-consistent
treatment of density and phase fluctuations, which yields an equation of state free of both infrared and
ultraviolet divergences. This theory is valid both for homogeneous and trapped systems in {\em all} dimensions, and
has been shown to be consistent with well-known results in the appropriate limits \cite{Low_D_Theory_2b}.
Furthermore, such an approach enables
a direct determination of quasi-condensate density profiles.

In the harmonic trap, one obtains
an appropriate finite temperature nonlinear Schr\"{o}dinger equation for the quasi-condensate, coupled
to the Bogoliubov-de-Gennes equations for the excitations. 
Solving these self-consistently  within the Thomas-Fermi approximation
yields a mean field approximation for
the quasi-condensate spatial extent. This is referred to as the
temperature-dependent Thomas-Fermi 
radius, and is denoted by  $R_{\rm TF}(T)$. 
This quantity depends on the quasi-condensate depletion $n^\prime(z)=n_{\rm T}(z)$, via
$R_{\rm TF}(T)= \sqrt{2\mu^{\prime} / m \omega_{z}^{2}}$
where  $\mu^\prime=\mu-2 \kappa n^\prime(0)$
is the `renormalized' chemical potential. Here
$\kappa$ is the effective 1D coupling constant which is fixed by the atom mass, $m$,
the 3D scattering length, $a_{\rm 3D}$, and the transverse confining frequency $\omega_{\perp}$
via $\kappa=a_{\rm 3D}/2\pi\ell_\perp^2$, 
where $l_{\perp}=\sqrt{\hbar/m \omega_{\perp}}$ is the harmonic oscillator length in the
transverse direction \cite{Olshanii}.
Density profiles are obtained from the above equations in the local density approximation,
with thermal profiles outside the quasi-condensate region obtained by solving the
equation of state for the normal gas.

The  first-order correlation function can be written as
$ g^{(1)}(0,z) =
{\rm exp}\left( -\langle\left[\hat{\chi}(z)-\hat{\chi}(0)\right]^2 \rangle /2 \right),$
where  $\hat{\chi}(z)$ is the operator for the phase, defined by expressing the Bose field operator
$\hat{\Psi}(z)$ in the density-phase representation as 
$\hat{\Psi}(z) = \sqrt{n(z)} {\rm exp}\{i \hat{\chi}(z)\}$.
In the theory of Andersen et al. \cite{Low_D_Theory_2a}, where density fluctuations are explicitely taken
into consideration, the above exponent assumes, in the purely one-dimensional geometry considered here, the 
following form
\begin{eqnarray}
\langle
\left[
\hat{\chi}(z)-\hat{\chi}(0)
\right]^2
\rangle & = & {4\pi \kappa l_z^4\over R_{\rm TF}^{3}(T)}
 \sum_{j=0}  2N(\hbar \omega_j) \\ \nonumber
& & \Bigg[
A_j^2\left(P_j(z/R_{\rm TF}(T))-P_j(0)\right)^2 \\ \nonumber
& & -B_j^2\left({P_j(z/R_{\rm TF}(T))\over 1-(z/R_{\rm TF}(T))^2}-P_j(0)\right)^2
\Bigg] \;.
\end{eqnarray}  
Here $P_j(z)$ are Legendre polynomials of order $j$,
with $A_j=\sqrt{(j+1/2)\mu^\prime/\hbar\omega_j}$, and
$B_j=(\sqrt{(j+1/2)\hbar\omega_j/\mu^\prime})/2$. The
frequencies are given by $\omega_j=\sqrt{j(j+1)/2}\;\omega_z$,
where $l_{z}$ is the harmonic oscillator length
corresponding to a longitudinal confining frequency $\omega_{z}$.
$N(\hbar \omega_{j})$ is the usual Bose distribution function.

The appearance of the temperature-dependent
quasi-condensate size $R_{\rm TF}(T)$ in the prefactor ensures that
density fluctuations are explicitly maintained in the above expression.
The general expression quoted above can be readily reduced to the
conventional theory  which ignores quasi-condensate depletion  \cite{Low_D_Theory_1,Low_D_Theory_3a}
by replacing $R_{\rm TF}(T)$ by the corresponding
zero temperature quasi-condensate size $R_{\rm TF}(0)$ at the same {\em total} atom number,
and ignoring the $B_j$ contributions.
Since $R_{\rm TF}(T) \propto \sqrt{\mu'}$, the former step is equivalent to replacing the renormalized
chemical potential at temperature $T$ by the corresponding zero-temperature one for the same total atom number.
Then, the `classical' approximation
$N(\hbar \omega_{j}) \approx k_B T / \hbar \omega_{j}$ leads to the definition of the characteristic temperature 
$T_{\phi}=(\hbar \omega_{z})^{2} N/k_{B} \mu$ \cite{Low_D_Theory_1}.

For more details on the implementation of this theory to trapped gases, the reader is referred to
\cite{Low_D_Theory_2a,Low_D_Theory_2b,Low_D_Theory_2c,Low_D_Theory_2d}.

\end{document}